\documentclass[twoside,a4paper]{article}
\usepackage{amsmath,graphicx,amssymb,fancyhdr,amsthm,,textcomp} 
\usepackage{soul} 

\usepackage{graphicx}
\usepackage{amsfonts}
\usepackage{amsmath}
\usepackage{rotating}
\usepackage{amssymb,amsthm}
\usepackage{xcolor}
\usepackage[normalem]{ulem}

\theoremstyle{definition} 
  
\theoremstyle{remark}  
  
\def\beq{\begin{eqnarray}}  
\def\eeq{\end{eqnarray}}  
\def\bsp{\begin{split}}  
\def\esp{\end{split}}

\def\Ep{\mathcal{E}}

 \newcommand{\po}{\Psi_0}
\newcommand{\pa}{\Psi_1}
\newcommand{\pb}{\Psi_2}
\newcommand{\pc}{\Psi_3}
\newcommand{\pd}{\Psi_4}

\begin{document}

\title{Geometric Horizons: A Frame Approach} 

\author{{\large\textbf{A. A. Coley$^{1}$ and D. D. McNutt$^{2}$   }}
\vspace{0.3cm} \\ 
$^{1}$Department of Mathematics and Statistics,\\
Dalhousie University,
Halifax, Nova Scotia, Canada B3H 3J5
\vspace{0.3cm} \\ 
$^{2}$ Faculty of Science and Technology,\\
University of Stavanger, 
N-4036 Stavanger, Norway  \\
 \vspace{0.3cm} \\
\texttt{aac@mathstat.dal.ca,david.d.mcnutt@uis.no}}  
\date{\today}  
\maketitle  
\pagestyle{fancy}  
\fancyhead{} 
\fancyhead[EC]{}  
\fancyhead[EL,OR]{\thepage}  
\fancyhead[OC]{}  
\fancyfoot{} 
  
\begin{abstract}

In the numerical investigation of the physical merger of two black holes, it is crucial to locate a black hole locally. This is usually done utilizing an apparent horizon. An alternative proposal is to identify a geometric horizon (GH), which is characterized by a surface in the spacetime on which the curvature tensor or its covariant derivatives are algebraically special. This necessitates the choice of a special null frame, which we shall refer to as an algebraically preferred null frame (APNF). The GH is then identified by surfaces of vanishing scalar curvature invariants but, unfortunately, these are difficult to compute. However, the algebraic nature of a GH means that the APNF plays a central role and suggests a null frame approach to characterizing the GH. Indeed, if we employ the Cartan-Karlhede algorithm to completely fix the null frame invariantly, then all of the remaining non vanishing components of the curvature tensor and its covariant derivatives are Cartan scalars. Hence the GH is characterized by the vanishing of certain Cartan scalars. 

A null frame approach is useful in the  numerical investigation of the merger of two black holes in general, but we will focus on the application to identifying a GH. We  begin with a review of the use of APNF and GH in previous work. The APNF is then defined and chosen so that the  Weyl tensor is algebraically special, and we must examine the covariant derivatives of the Weyl tensor in this frame. 
 We show how to invariantly fix the null frame, and hence characterize the APNF, and describe how to then identify the GH using the zero-set of certain  Cartan scalars. Our ultimate aim is to apply this frame formalism to the numerical collapse of two black holes. As an example, we investigate the axisymmetric evolution of a two black hole Kastor-Traschen spacetime.
\end{abstract}

%
%
%

\newpage

\section{Introduction}

The event horizon of a black hole ({{BH}}) solution in General Relativity (GR) is defined as the boundary of the non-empty complement of the causal past of future null infinity; i.e., the region for which signals sent from the interior will never escape. The event horizon is a teleological surface, as it can only be located with knowledge of the entire development of the spacetime \cite{AshtekarKrishnan}. To examine the interaction of realistic BH with their environment in numerical GR \cite{T}, in the 3+1 approach or in the Cauchy-problem in GR, the location of a BH must be determined locally \cite{AnderssonMarsSimon,Jaramillo}, and any such localization may not rely solely on the existence of the event horizon, as BH are expected to
undergo evolutionary processes and are typically dynamical.

The concept of closed trapped surfaces without border, which are compact spacelike surfaces such that the expansions of the
future-pointing null normal vectors are negative, was proposed in \cite{Penrose:1964wq}. Consequently, for time-dependent  BHs,  apparent horizons (AHs) are used instead of event horizons. AHs are quasi-local surfaces defined as the zero-set of the vanishing expansion of a null geodesic congruence normal to a trapped surface with spherical topology. A
related concept to trapping surfaces are are marginally trapped surfaces
(MTSs) which are two-dimensional (2D) surfaces for which
the expansion of the outgoing null vector normal to the
surfaces vanishes and  marginally outer trapped surfaces
(MOTSs) which MTSs with the additional condition that the expansion of the ingoing null vector normal to the
surfaces is negative. Assuming a smooth time evolution for the MOTSs,
the 2D surfaces can be combined to construct a three-dimensional (3D) surface
known as a marginally trapped tube (MTTs) \cite{BoothFairhurst}. If the MTT is foliated by MOTSs and the expansion of the ingoing null vector normal to the surface is negative then it is called a dynamical horizon ({{DH}}) \cite{AshtekarKrishnan}. Lastly, if the Lie derivative of the outgoing expansion in the direction of the ingoing null vector is negative, the DH is a future outer trapping horizon (FOTH). Unlike the event horizon, the AH and MTTs are quasi-local,
and they are intrinsically foliation-dependent, and hence observer dependent so that different observers may observe different MTTs, and 
trapping horizons and DHs can be
non-unique \cite{AG2005}.  
A DH is particularly well-suited to analyze realistic dynamical processes involving BH
such as BH growth and coalescence \cite{AshtekarKrishnan}. 
This
motivates the idea of using MTTs 
as viable replacements 
for the event horizon
of BH \cite{Senov}.  In this paper we will adopt the common usage of calling the FOTH an AH, while clearly noting that this surface is distinct from the AH defined using trapping surfaces.

In a bit more detail, we wish to study the location of an appropriate horizon of a BH in four dimensions (4D). Let $\ell, n$ be two independent 
future directed null vectors, and let $\theta_{(\ell)}, \theta_{(n)}$ be their corresponding expansions.
There are various notions of horizon, many of which depend on a surface on which the expansion of an appropriate null congruence is zero
\cite{AshtekarKrishnan}, including a trapping surface ($\theta_{(\ell)}<0, \theta_{(n)}<0$), a MOTS ($\theta_{(\ell)} = 0, \theta_{(n)}<0$), an AH which is a future outermost MTTs,
and a DH which is foliated by MTSs. These surfaces depend explicitly on the choice of  $\ell, n$ (and strictly speaking are orthogonal to them and of spherical topology), and they are also
foliation dependent \cite{AG2005}. That is, a smooth 3D spacelike submanifold  of a spacetime is said to be a DH if it can
be foliated by closed 2D submanifolds such that
$\theta_{(\ell)}=0, \theta_{(n)}<0$. 
\newpage

An alternative proposal to identify the boundary of a BH is provided by the  geometric horizon ({{GH}}) conjectures \cite{CMS}. In these conjectures, the horizon is identified by surfaces in the spacetime on which the curvature tensor or its covariant derivatives
are algebraically special, in that positive boost weight (b.w.) terms in an appropriate null frame are zero \cite{class}. That is, for a BH spacetime arising in the dynamical collapse or merger of real BH, the geometry is typically of general algebraic type away from the horizon, but is more algebraically special on the horizon \cite{CMS}. 

In 4D, we may always choose the null frame relative to which either the Weyl tensor or the Ricci tensor is of algebraic or Petrov type {\bf I}; we shall refer to this an an algebraically preferred null frame (or APNF). The GH is then identified by a surface on which the spacetime is  more algebraically special; i.e., where  positive b.w. components vanish on this surface. Often, the vanishing of b.w. +1 components can be characterized by the vanishing of scalar polynomial curvature invariants (SPIs). If the spacetime is of type {\bf II} or {\bf D}, the covariant derivative of the Weyl or Ricci tensor can be used instead to define the GH, as one of these tensors will, in general, be of type {\bf I}.

In principle, the GH could be identified by surfaces of vanishing SPI  \cite{CMS} but, unfortunately, the relevant SPIs are difficult to calculate. However, the algebraic nature of a GH means that the APNF plays a central role, and also suggests a null tetrad/frame approach. In this approach we will first choose a null tetrad (i.e., $\ell$ and $n$, and a complex null vector $m$ and its complex conjugate $\bar{m}$) such that the b.w. +2 terms of the Weyl tensor (or Ricci tensor) are zero. This will ensure that the algebraic type is manifestly type {\bf I}. If  the Weyl tensor is of type {\bf II}, we will choose a frame relative to which its covariant derivative is of type {\bf I}, which ensures that the null frame is an APNF. One of the main problems of the usual foliation dependent  approach is the need for a "sound" criterion for selecting a preferred $\ell$ and $n$ (and hence the AH or DH \cite{JMMS}); the APNF approach addresses this problem according to algebraic specialization.

Relative to the APNF, the curvature tensor or its covariant derivatives are of algebraic type {\bf I} outside of the horizon, but they  will become more algebraically special on the horizon, thereby identifying the GH. If we now employ the Cartan-Karlhede algorithm to completely fix the frame invariantly \cite{GANG}, then all of the remaining non vanishing components of the curvature tensor and its covariant derivatives are Cartan scalars and these quantities are easier to compute than their related SPIs. Consequently, we identify a GH by the vanishing of, for example, positive b.w.
$C_{abcd;e}$  terms via Cartan scalars
\cite{GANG}, which can be computed in a straightforward manner. In particular, the vanishing positive b.w. $C_{abcd;e}$  terms can be related to the Newman-Penrose (NP) spin coefficients via the NP formalism, and so the GH can also be identified by the vanishing of particular spin coefficients. Notably, the spin coefficients $\rho$ and $\mu$ can often be expressed in terms of Cartan invariants and are directly related to the corresponding expansions of the null directions of the APNF; i.e., $\theta_{(\ell)} \propto \rho$ and $\theta_{(n)} \propto \mu$. We will focus on the vanishing of these spin-coefficients in order to define the GH.  These conditions are similar to the conditions identifying DH and AH in the normal approach,  but they are now Cartan scalars with a specific geometric and algebraic interpretation.

\subsubsection{Review of previous work}

In many cases it has been shown that relative to  an APNF the GH is identified by
the vanishing of $\rho, \mu$. For example: I. For stationary spacetimes
with stationary horizons (such as the Kerr-Newman-NUT-anti de Sitter metric) the spacetime is everywhere of Weyl and Ricci type {\bf D}, and the APNF is adopted so that both these algebraic conditions are manifest, and the location of the GH is then obtained by the surface on which the covariant derivative of the Weyl and Ricci tensors are of type {\bf II}, which is equivalent to the scalar $\rho$ vanishing. In fact, this property is realized for a class of horizons that generalize the Killing horizons, known as weakly isolated horizons (WIHs)  \cite{AshtekarKrishnan}, which will be discussed later in the paper. II.  In the case of spherically symmetric dynamical BHs, the Weyl tensor is of type {\bf D}, and for vacuum solutions or known exact solutions such as Vaidya or Lemaitre-Tolman-Bondi (LTB) dust solutions (in which the horizons are known and are, in fact, isolated horizons (IH)) the Ricci tensor is simple and does not help to identify the GH \cite{CMS}. So we choose the APNF adapted to Weyl type {\bf D}, in which case the type {\bf II}  GH surface of the covariant derivative of the Weyl tensor is identified by $\rho =0$. III. In the case of the non-spherically symmetric
quasi-spherical Szekeres dust models, the spacetimes are of 
Weyl {\bf D} and Ricci type {\bf I}, and the APNF is adapted to this. The covariant derivative of the Weyl tensor is then considered, and the GH is shown to be identified by $\rho =0$ \cite{szek}. IV. In the case of the 4D Kastor-Traschen multi-N-BH spacetimes, the  Weyl tensor is of type {\bf I} (for $N>1$). However, the Ricci tensor is of type {\bf D}, and the APNF
is adapted accordingly. The surface on which the covariant derivative of the 
Ricci tensor is of type {\bf D}, for which $\rho=0, \mu=0$, then identifies the GH \cite{CMKT}. We discuss these examples in more detail later.

Motivated by these observations, we will consider vacuum spacetimes in which the Ricci tensor is identically zero, perhaps representing the physically interesting situation of the merger of two BH. Here, the APNF will be chosen so that the  Weyl tensor is of type {\bf I} and we must examine the covariant derivative of the Weyl tensor in this frame. We will motivate the choice of the APNF and the use of the spin coefficient $\rho$ relative to this frame to determine the GH using the NP formalism. We will also consider the non-vacuum 2-equal-mass Kastor-Traschen multi-black hole solution and derive an alternative APNF based on the Weyl tensor.

\subsection{The Geometric Horizon  Conjectures}

In \cite{CMS} a foliation invariant and more geometrical approach, in which the curvature
tensor (when treating the Weyl and Ricci tensor as curvature operators \cite{BIVECTOR}) 
is algebraically special relative to the alignment classification (using the boost weight decomposition) \cite{class}
was proposed to identify the BH horizon.  Using recent results in invariant theory, such
geometric BH horizons can be identified by the alignment
type {\bf II} or {\bf D} discriminant conditions in terms of
scalar curvature invariants, which are not dependent on spacetime
foliations.

As a brief review of the alignment classification, consider the effect of a boost on a given null coframe $\{  n, \ell, \bar{m}, m \}$,
\beq
(\ell, n) \to (e^{\lambda} \ell, e^{-\lambda} n), \label{badboost}
\eeq
\noindent where $\lambda$ is real-valued.
This gives rise to the concept of a  {\it boost weight} $b \in \mathbb{Z}$ such that for an arbitrary component of a rank $r$ tensor ${\bf \hat{T}}$ with respect to the null coframe, a boost in the plane, $(\ell, n)$, gives the transformation
\beq \hat{T}_{a_1 ... a_r} \to e^{b\lambda} \hat{T}_{a_1 ... a_r},  \eeq
\noindent where the indices $a_i, \ldots a_r$ range from $1$ to $4$, and the integer $b$ is the boost weight vector of the component $\hat{T}_{a_1 ... a_r}$ which records the difference in the number of appearances of $\ell$ and $n$, respectively, in the associated tensor product of a given component $\hat{T}_{a_1 ... a_r}$. We can write the tensor ${\bf \hat{T}}$ in the following decomposition:
\beq
{\bf \hat{T}} = \sum\limits_{b} ({\bf \hat{T}})_{b}.
\eeq
\noindent Here $({\bf \hat{T}})_{b}$ denotes the projection onto the subspace of components of boost weight $b$.

Defining the maximum boost weight of a tensor, ${\bf \hat{T}}$, for a null direction $\ell$ as the boost order, we denote this as $\mathcal{B}_{{\bf \hat{T}}}(\ell)$. For a given null direction $\ell$, $\mathcal{B}_{{\bf \hat{T}}} (\ell)$ remains unchanged under boosts, spatial rotations and null rotations about $\ell$, implying that the choice of $n$ will not affect the integer value of $\mathcal{B}_{{\bf \hat{T}}} (\ell)$. This implies the definition is dependent on the choice of $\ell$. Defining $B_{{\bf \hat{T}}}$ as the maximum value of $\mathcal{B}_{{\bf \hat{T}}} (\ell)$ over all possible choices of $\ell$, the existence of a $\ell$ with $\mathcal{B}_{{\bf \hat{T}}} (\ell) <B_{{\bf \hat{T}}}$ is an invariant property of the tensor ${\bf \hat{T}}$. We will say $\ell$ is ${\bf \hat{T}}$-aligned if $\mathcal{B}_{{\bf \hat{T}}}(\ell) < B_{{\bf \hat{T}}}$.

The Weyl tensor, or any rank two tensor, ${\bf T}$, in arbitrary dimensions can be broadly classified into six {\it alignment types}: if for all null directions $\ell$, $\mathcal{B}_{{\bf T}} (\ell)=2$ then ${\bf T}$ (or the Weyl tensor) is of alignment type $G$, if there exists an $\ell$ such that $\mathcal{B}_{{\bf T}} (\ell) = 1, 0,-1,-2$, then ${\bf T}$ (or the Weyl tensor) is of alignment type $I, II, III,$ or $N$, respectively, while if {\bf T} (or the Weyl tensor) vanishes then it belongs to alignment type $O$. In addition, if two null directions can be determined which are not proportional and each give rise to alignment type {\bf II}, then ${\bf T}$ (or the Weyl tensor) is of special alignment type ${\bf D}$. For higher rank tensors, like the covariant derivatives of the Weyl tensor or Ricci tensor, the alignment types are still applicable despite the possibility that $|\mathcal{B}_{{\bf T}} (\ell)|$ may be greater than two.

Therefore, a particular set of SPIs and Cartan invariants vanish on the GH \cite{CMS} due to the fact that on the horizon the curvature tensor and its covariant derivatives must be of type {\bf II}/{\bf D} relative to the alignment classification \cite{class}. We note that in 4D, there is no type {\bf G} for the Weyl or Ricci tensor and so algebraically general here means type {\bf I}.

\paragraph{GH Conjecture I:}
{\em{If a BH spacetime is zeroth-order algebraically general, then on the GH the spacetime
is algebraically special. We can identify this GH using scalar curvature invariants.}}

\paragraph{Comments:}      
The algebraic conditions expressed in terms of $SPI$s essentially define a {\em{GH}}.
We also note that $SPI$s may not specify the GH completely in the sense that the invariants may vanish at particular points such as the fixed points of an isometry or along an axis of symmetry. Unlike 
AHs, a GH does not depend on a chosen foliation in the
spacetime.

We are primarily interested in applications in 4D, and particularly
in numerical computations.
In physically relevant problems with dynamical evolution, such as
asymmetric collapse and BH coalescences, the 
horizon might not be unique,  or may not be defined by the specified invariants at all, and the conjectures may have to be modified accordingly
(e.g., it may be appropriate to replace the vanishing of invariants in the definition of a 
GH as an algebraically special hypersurface, with the conditions that the magnitudes of certain $SPI$s take their smallest values). Furthermore, it may be necessary to consider specializations of the positive b.w. terms that do not fit into the standard broad alignment types.

 On the horizon, the Riemann tensor and its covariant derivatives will always be more algebraically special than the surrounding regions of spacetime. If the Riemann tensor is already algebraically special (i.e., of type {\bf II}/{\bf D}) then the Ricci and Weyl tensor in the same frame will be algebraically special as well.

\paragraph{GH Conjecture II:}
{\em{If the whole spacetime is zeroth-order algebraically special (and on the horizon the spacetime is thus also algebraically special) and if the  whole spacetime has an algebraically general first order covariant derivative of the Riemann tensor, $R_{abcd;e}$, then on the horizon $R_{abcd;e}$ will be  algebraically special and we can identify this surface using SPIs.}} 
\vspace{3 mm}

\subsubsection{Scalar polynomial (curvature) invariants }

As a preliminary tool to verify the standard algebraic types, we can use discriminants to determine the necessary conditions in terms of simple SPIs.
BH spacetimes are completely characterized 
by their $SPIs$ \cite{CH}. The necessary type {\bf  II}/{\bf D} discriminant condition ${^4}D_4=0$, where ${^4}D_4$ is a SPI in terms of the trace-free symmetric Ricci tensor $S$ in 4D  given in \cite{CH}.

The necessary real conditions for the Weyl tensor to be of type {\bf  II}/{\bf D} were also given 
explicitly in \cite{CH}:
these 2 real conditions are equivalent to the real and imaginary parts of the complex syzygy ${\mathcal{D}} \equiv I^3-27J^2=0$ in terms of
the complex Weyl tensor, where the $\Psi_i$ are components of the complex-valued Weyl spinor in the NP formalism \cite{kramer}, with $I$ and $J$ defined as
\beq I = \pd \po - 4 \pc \pa + 3 \pb^2, \label{Iinv} \eeq

\beq J = det \left| \begin{array}{ccc} \pd & \pc & \pb \\ \pc & \pb & \pa \\ \pb & \pa & \po \end{array} \right|. \label{Jinv} \eeq

The NP formalism is also useful for computing the expansion of the null frame vectors,  
which is related to the real part of the NP spin coefficient $\rho$. Algebraic specialization of the curvature tensors and their covariant derivatives
can also be established explicitly by chosing 
future directed null vectors  $\ell, n$ and consequently an APNF  
such that the vanishing of positive b.w. terms can often be related, by the NP formalism, to the Cartan invariants $\rho, \mu$, such that in the  APNF 
the GH is then determined by $\rho=0$ or $\mu = 0$.

Useful {\it necessary} conditions can be obtained by
considering just one of these real type {\bf II/D} constraints (or just the real or imaginary
part of $\mathcal{D}=0$). More practical {\it necessary} conditions can be obtained by
considering the trace-free symmetric operator
$C_{abcd}C^{ebcd} - 2 W_2 \delta_a^{~e}$ where $W_2 = C_{abcd} C^{abcd}$ \cite{CH}. To consider whether the
covariant derivatives of the Weyl or Ricci tensors  are
of type {\bf II} or  {\bf D}, we can use the eigenvalue structure of the operators associated with the derivatives of the Weyl or Ricci curvature.

\newpage

\subsection{Numerics and Discussion}

In order to investigate the applicability of these definitions and conjectures, it is important to study the existence of GHs in physically relevant situations. This suggests the use of numerical descriptions in 4D of physical asymmetric collapse or BH coalescences. This is due to the fact that in numerical relativity simulations involving BHs, the locations and properties of the BHs are tracked using AHs or trapping horizons \cite{Booth2005}.  For example, AHs are utilized in  the generation of waveforms for gravitational waves generated from BH formation through merger or stellar collapse in numerical relativity. More recently, the observations by the LIGO collaboration of gravitational waves from BH mergers has relied upon the comparison with templates obtained using numerical simulations based on AHs \cite{LIGO}.  The study of AHs is an active field of research and it has become clear that MOTSs are better behaved numerically than previously expected \cite{Schnetter:2006yt}.

If the results of the conjectures are to be useful to numerical relativists,  the computability of GHs must be considered. Generally, the Cartan invariants are less computationally expensive to compute than the related SPIs \cite{GANG}. We note that if we wished to investigate the existence of GHs in a numerical solution once the numerical calculation is complete, then directly employing the Cartan invariants would be difficult, as these are frame dependent quantities and rely on the choice of a specific frame. However, we could use the same frame used in numerical simulations (which is not an APNF) once the Cartan-Karlhede algorithm has been implemented and relate the resulting frame to the frame used in numerical simulations. This approach has been partially used in numerical relativity where particular wave extraction techniques can be classified using the Weyl scalars from the NP formalism in 4D. However, in order to employ the NP approach fully we must completely fix the frame so that the resulting ``Weyl scalars'' are Cartan scalars.

A simple approach to numerical simulations of BH binaries (BBHs) is to model the BHs as punctures in the fabric of spacetime \cite{PhysRevLett.78.3606} and the locations of the BH punctures are tracked with AHs \cite{EvansFerguson}. Employing the Einstein toolkit infrastructure \cite{Loffler:2011ay} for such a calculation, the location of MOTSs is determined numerically using a horizon finder
(which is based on the eigenvalues of the  stability operator for MOTS \cite{AnderssonMarsSimon,Andersson_2009}). Such a horizon finder is capable of finding MOTSs, even when they are highly distorted \cite{Pook-Kolb} (see also \cite{Pook}).

We are particularly interested in the application to numerical studies in vacuum BBH spacetimes and especially, at least in the first instance, in the axisymmetric case. Non-vacuum BBH spacetimes, such as in the Kastor-Traschen (KT) 2 BH spacetimes \cite{KT} (see later), are also of interest. In this direction, the head-on collisions of two unequal mass non-spinning BHs starting with Brill-Lindquist initial data, representing a BBH system at a moment of time-symmetry, is of particular interest for future work.

\newpage

\section{Examples}

There are many examples  \cite{GANG} that 
support the GH conjectures. 
There is also motivation for the conjectures
from analytical results
\cite{AshtekarKrishnan}.
For example, in 4D, and assuming the dominant energy condition, it was demonstrated that on the non-expanding WIH 
the Ricci and Weyl tensors are of type {\bf II/D} \cite{Ashtekar}
and that the covariant derivatives of the Riemann tensor 
on a WIH are of type {\bf II} on the horizon \cite{CMS}.

We note that when a star collapses to
form a BH, the exterior of the BH eventually settles down to a
stationary state, most likely described by the ``unique'' Kerr metric (which is of
type {\bf D}). Regardless of what the inner region of the
BH settles down to, this leads by continuity to the
expectation that  the interior near the horizon should contain a region that is reasonably close to the inner region of the Kerr metric. Within the BH event horizon, the Kerr metric admits a null inner horizon.
There are a variety of analytic arguments, mathematical results, and numerical
simulations that indicate that this surface maintains the inner horizon's
character as a null surface \cite{AshtekarKrishnan,Ashtekar}. 
This supports the notion that the horizon is smooth and unique at later times
and, in principle, can be identified by algebraic/geometrical conditions. It is possible that as
we follow this unique, smooth surface back in time (during the physics of collapse or merger),
this surface suffers a bifurcation and this surface is no longer unique or
smooth (or even differentiable). But it is plausible that there exists a unique, smooth
GH that shields all other horizons.

The GH conjecture is intended to apply to {DHs}, which are     
much more difficult to study.
There are examples of dynamical BH solutions that admit GH
\cite{CMS}, such as dynamical BH solutions which are conformally
related to a stationary BH solutions and the imploding spherically
symmetric metrics [17]. For any dynamical spherically symmetric metric a scalar
invariant  will detect the unique, invariantly defined dynamical
horizon.
For the spherically symmetric dynamical BH and dynamical black
holes conformally related to stationary BH the GHs correspond to MTTs.
However, in general a GH will not be a MTT, as the preferred null direction
will not necessarily be geodesic and surface forming. 

 Even though the discriminant approach will provide necessary conditions for the Riemann tensor or its covariant derivatives to be algebraically special, this approach is difficult to implement practically. Instead, the frame approach provides a more direct approach to verify the type {\bf II}/{\bf D} property of the horizon by constructing the required frame where the Ricci tensor or Weyl tensor become algebraically special on the GH. To determine this frame, we can use the Cartan-Karlhede algorithm \cite{GANG}.  For a BH solution, in general, an APNF can be determined through the condition that the Riemann tensor or its covariant derivatives are of algebraic type {\bf I} outside of the horizon. We anticipate that these tensors  will become type {\bf II} or more algebraically special on the horizon, relative to this APNF.

\newpage

\subsection{I: 4D Kerr-Newman-NUT-(Anti) de Sitter metric}

First we consider the Kerr-Newman-NUT-(Anti)-de Sitter solution which admits stationary horizons \cite{GANG}; the metric is given in \cite{PlebDem1976}.
The horizon locations coincide with the roots of the polynomial, $Q(r)$, appearing in the metric. In this example, the coordinate expressions for the discriminant SPIs are very large and are not easily factored. In comparison, the coordinate expressions for the Cartan invariants are much easier to combine to construct simpler horizon detectors. Indeed, it can be shown that the covariant derivative of the Ricci and Weyl tensors are of algebraic type {\bf II/D} on the horizon using the spinor formalism.

We define an appropriate APNF in which
the non-trivial NP curvature scalars are $\Lambda_{NP} = \frac16 \Lambda$, 
$\Psi_{2}$ and $\Phi_{11}$ (in which the Weyl and Ricci tensors are explicitly algebraically special) \cite{GANG}.
The Cartan invariants arising from the Riemann tensor are the real and imaginary parts of $\Psi_2$, which are functionally independent, along with $\Phi_{11}$ and $\Lambda_{NP}$, as functionally dependent invariants. The isotropy group of the Riemann tensor is generated by boosts and spins. The frame is completely fixed, up to spins, using the covariant derivative of the Riemann tensor and we can compute the non-zero components of the  covariant derivatives of $\Psi$ and $\Phi$. Computing the spin-coefficient $\rho = \mu$, it can be shown that these are Cartan invariants and take the form \cite{GANG}:

\beq \rho = \mu = - \frac12  \frac{\sqrt{Q} [r + i(a\cos\theta + l)]}{\tilde{\rho} | a\cos\theta + l + i r|^2 }.  \eeq

The horizons' locations are now easily determined as the roots of $Q(r)$. From this property, all b.w. -1 and +1 terms are zero when evaluated on the horizon. This implies that the first covariant derivative of the Weyl and Ricci tensors are of type {\bf D}. The condition that
$C_{abcd;e}$ is of type {\bf II/D} on the horizon
in the APNF is achieved here by $\rho = 0$.

\subsection{II: Imploding Spherically Symmetric Metric}

As another example, we introduce the imploding spherically symmetric spacetime in advanced coordinates:

\beq ds^2 = - e^{2\beta(v,r)} \left(1 - \frac{2m(v,r)}{r}\right) dv^2 + 2 e^{\beta(v,r)}dv dr + r^2 d \Omega^2, \label{SphSymMtrc} \eeq

\noindent where $m(v,r)$ is the mass function and $\beta(v,r)$ is an arbitrary function. We have exhausted all gauge freedom, so that, in general, no further simplification of the Einstein tensor is available.

This metric is not an exact solution per se. However, it will contain exact solutions once $m$ and $\beta$ are specified. As an example, if $\beta$ vanishes and $m$ is a function of $r$ alone, this yields the imploding Vaidya solution (which provides explicit examples of a transition from a DH to an WIH for particular choices of these metric functions), and the unique dust solution with $\beta_{,v} = 0$ is given by the LTB solution in which
the frame vectors on the horizon are geodesic since $\beta_{,v} =0$ (and hence also admits an isolated horizon). 
In dynamical spherically symmetric BHs the GH corresponds to a MTT \cite{CMS}; however, for more general BHs the GH will not be a MTT, since the preferred null direction need not be geodesic and surface forming.

The following vectors are chosen as the future pointing null geodesic vector fields \cite{CMS}:

\beq \ell = \partial_v + \frac12 \left( 1 - \frac{2m}{r}\right) \partial_r ,~~n = e^{-\beta} \partial_r,  \label{SSln} \eeq
\noindent where $\ell_a n^a = -1$, and complete the non-coordinate basis using the complex spatial vector and its complex conjugate:
\beq m = \frac{1}{\sqrt{2}r} \partial_\theta + \frac{i}{\sqrt{2} r \sin(\theta) } \partial_\phi.  \label{SSmmb} \eeq
\noindent Relative to this null frame basis the metric is now diagonalized: 
$g_{ab} = -\ell_{(a} n_{b)} + m_{(a} \bar{m}_{b)}$,
and the future null expansions are thus:
\beq \theta_{(\ell)} = \frac{e^{\beta}}{r}\left(1-\frac{2m}{r}\right),~~~ \theta_{(n)} = - \frac{2e^{-\beta}}{r}. \eeq

The unique spherically symmetric surface $r-2m(v,r) = 0$, which is in fact a FOTH  \cite{Faraoni2017}, is characterized by $\theta_{(\ell)} = 0$ \cite{Senov}, and this surface will be denoted as $\tilde{\mathcal{H}}$. Depending on the nature of the normal vector, $n_a$, to $\tilde{\mathcal{H}}$, this surface can be timelike, null or spacelike.  In the case that $m_{,v} \neq 0$, $\tilde{\mathcal{H}}$ will be spacelike and hence a DH or timelike and hence a timelike membrane, while the vanishing of  $m_{,v}$ on  $\tilde{\mathcal{H}}$ implies that the surface $\tilde{\mathcal{H}}$ is an isolated horizon \cite{Senov}. If instead we consider the class of spherically symmetric metrics  for which $m_{,v}=0$, the condition for $\ell^a$ and $n^a$ to be geodesic requires that $\beta_{,v} = 0$ and it follows from the NP formalism that the Weyl and Ricci tensors and their covariant derivatives are of algebraic type {\bf II/D} on the null horizon $\hat{\mathcal{H}}$ \cite{CMS}, even if these spacetimes are not exact solutions to the EFE.

Let us consider the general case with no further constraints from the EFE (i.e., this will not represent an exact solution) when $\hat{\mathcal{H}}$ is spacelike. We can apply the NP formalism to show that  relative to the chosen frame, the covariant derivative of the Riemann tensor is no longer of type {\bf II} on $\tilde{\mathcal{H}}$; instead, the structure of the covariant derivative of the Riemann tensor changes in a predictable manner on $\tilde{\mathcal{H}}$. Furthermore, the change in the structure of the covariant derivative of the Riemann tensor can be characterized in a frame-independent manner by the vanishing of a SPI.

\subsubsection{APNF and spin coefficients}

 Using the coframe given by eqns. \eqref{SSln} and \eqref{SSmmb}, the sole non-zero component of the Weyl spinor is $\Psi_2$, and so $\ell$ and $n$ can be used to construct the APNF. The boost parameter can be fixed using the Ricci tensor since the only non-zero NP curvature scalars for the Ricci spinor are $\Phi_{00}, \Phi_{11}, \Phi_{22}$ and the Ricci scalar $R$.  We can relate the components of the Ricci tensor to those of the Ricci spinor using $2\Phi_{00} =  R_{11}, ~~2 \Phi_{22} =  R_{22}, 4 \Phi_{11} = (R_{12} + R_{34})$. All other components of the Ricci tensor vanish. The Cartan scalars $\Phi_{00}, \Phi_{11}$ and $\Phi_{22}$ have b.w. $+2$,0,$-2$, respectively. 

The non-zero Weyl spinor component is related to the only algebraically independent component of the Weyl tensor:
\beq \Psi_2 = - C_{1324}. \nonumber \eeq
Thus the Weyl tensor is of algebraic type {\bf D}, and the Ricci tensor is generally of algebraic type {\bf I} ($\Phi_{00} \neq 0$) relative to the alignment classification \cite{class}, and the above frame is an APNF. The isotropy group of the Riemann tensor and its covariant derivatives consists of spins.  The discriminant SPIs associated with the Weyl tensor and the Ricci tensor cannot detect the surface $\tilde{\mathcal{H}}$ defined by $r=2m(v,r)$; in order to construct an SPI that detects this surface we must use the covariant derivative of the Weyl tensor.

Using the covariant derivative of the Weyl and Ricci tensors and applying the differential Bianchi identities, the components can be expressed in terms of $\Psi_2$, $\Phi_{00} = \Phi_{22}$,$\Phi_{11}$, $\Delta \Phi_{11}$ and the spin coefficients $ \epsilon, \mu$ and
\beq  \rho = -\frac{e^\beta(r-2m)}{2r^2}. \label{SSrhomu} \eeq
In this APNF, $\rho = 0$ on the GH regardless of the EFE.

The covariant derivative of the Weyl tensor will have components of b.w. $+1, 0$ and $-1$. In particular, the non-zero components of b.w. $+1$ are \cite{CMS}: 

\beq C_{1214;3} = C_{1434;3} = C_{1213;4} = C_{1334;4} = 3 \rho \Psi_2, \label{SSBW10} \eeq
and $2C_{1423;1} = C_{1212;1} = C_{3434;1}  \neq 0$.

Interestingly, the components in \eqref{SSBW10} vanish on $\hat{\mathcal{H}}$ and hence identify the GH. That is, this subset of algebraically special b.w. $+1$ components identify the horizon. However, the remaining positive b.w. components do not
necessarily vanish.  Thus, while the covariant derivative of the Weyl tensor is not necessarily of type {\bf II} on $\tilde{\mathcal{H}}$, the vanishing of these components implies that this tensor is algebraically special on $\tilde{\mathcal{H}}$. However, the existence of $\hat{\mathcal{H}}$ clearly affects the structure of this tensor, and this condition on the structure of the Weyl tensor is reflected in the vanishing of a SPI. That is, 
for any spherically symmetric metric, the structure of the covariant derivative of the Weyl tensor changes on  $\hat{\mathcal{H}}$ where $r=2m(v,r)$ and this can be detected by a Weyl invariant $ I_C$ proportional to  $\rho \mu \Psi_2^4$, which vanishes on
$\tilde{\mathcal{H}}$.

This is to be expected since $\rho$ is a Cartan invariant  relative to the APNF which vanishes on $\hat{\mathcal{H}}$. The behaviour of the b.w. $+1$ components of the first covariant derivative of the Weyl tensor and the positive b.w. components of the second covariant derivative of the Weyl tensor  exhibit the same behaviour, where only a subset of the  highest b.w. terms vanish on $\tilde{\mathcal{H}}$, and that these components can be expressed in terms of $\rho$.

In general $m_{,v}$ is not necessarily zero on $\tilde{\mathcal{H}}$. However, additional constraints (EFE) will be imposed for physically realistic exact solutions (such as the Vaidya and LTB exact solutions alluded to above).  To study the subset of imploding spherically symmetric metrics which satisfy the EFE, we can impose conditions on the components of the Einstein tensor relative to the coframe basis, $\{ n_a, \ell_a, \bar{m}_a, m_a \}$ above. For example,
in the case of null or non-null radiation, $\Phi_{00} = 0$, which implies that the surface $r = 2m$ is null; i.e., $m_{,v} = 0$. In the case that $\Phi_{00} =0$, the Ricci tensor is of type {\bf II}.

As another example of the behaviour of an imploding spherically symmetric metric satisfying the EFE, we consider a perfect fluid solution and explicitly exclude the case of dust  (i.e., the exact LTB solution studied previously). The condition for a perfect fluid implies that the scalar, $I_{pf} \equiv \Phi_{00}\Phi_{22} - 4\Phi_{11}^2$, which can be explicitly computed in local coordinates, must vanish.  

Let us consider a subclass of perfect fluid solutions in which $\beta_{,v} = 0$ on the horizon. Evaluating $I_{pf} = 0$ on the surface $\tilde{\mathcal{H}}$ where $r=2m$ gives an expression for $m_{,v}$ on this surface. By explicitly calculating  $C_{1212;1}$ evaluated on $\hat{\mathcal{H}}$ (with $\beta_{,v} = 0$) we can represent this expression as \cite{CMS}: 

\beq C_{1212;1}~|_{r=2m} = \{~\}M + \{~\}M^2 + \{~\}M_{,v}, \nonumber \eeq
\noindent where we have omitted the long coordinate expressions in the curly brackets and $M \equiv{(m_{,v})^\frac12 }$. Thus
if $M, M_{,v} =0$  on $r=2m$, it follows that $C_{1212;1} = 0$ there. 
Computing $m_{,v} = f(v,r)$ 
in a neighbourhood of $r=2m$ (assuming that it is sufficiently smooth that appropriate expansions can be made), we obtain

\beq C_{1212;1} \sim (r-2m)^2 + 0 [(r-2m)^3]  \sim \rho^2 + 0 [\rho^3]. \eeq

Therefore all positive b.w. terms are zero when $\rho = 0$, and
we can conclude that for a well-behaved perfect fluid spherically symmetric solution,  in a neighbourhood of a point on the horizon, locally all of the b.w. $+1$ terms of $C_{abcd;e}$ are zero on the horizon. It remains to study the general spherically symmetric case for appropriate physically motivated matter tensors.

\newpage

\subsection{III: Quasi-spherical Szekeres dust models}

The quasi-spherical Szekeres dust (QSSD) solutions are a non-spherical generalization of the LTB dust models and can be considered as cosmological models and are potentially models for the formation of primordial BHs (PBH) in the early universe \cite{PlebDem1976}. 
Indeed, any collapsing QSSD solution where an AH covers all shell-crossings that will occur can be  considered as a model for the formation of a BH. 
As a BH solution, the QSSD models require extensive fine-tuning of the BH's mass and collapse time in order to avoid shell-crossings forming outside of the AH.  A subset of the collapsing QSSD solutions can thus describe the formation
of PBHs in the early universe, and perhaps have a more natural interpretation than their spherically symmetric counterparts  \cite{PlebDem1976}, if
the BH mass is within a small enough range.

The QSSD models are known to admit an AH at $R=2M$. It was shown in \cite{szek} that this hypersurface is, in fact, a GH \cite{CMS}. To do so a frame approach  was employed to compute the appropriate Cartan invariants arising from the Cartan-Karlhede algorithm \cite{CMS} in order to determine the existence of the GH. Indeed, in an APNF
the AH was detected by the vanishing of a Cartan invariant;
$\rho = 0$ (or $\mu=0$) provides a putative characterization for the GH \cite{szek}.

We write the metric for the $\beta' \neq 0$  QSSD solutions with vanishing cosmological constant as \cite{hellaby1996null}:  

\beq ds^2 = - dt^2 + \frac{\Ep^2 {Y'}^2}{1+2E} dz^2 + Y^2 [dx^2 + dy^2], \label{mtrc} \eeq

\noindent where $Y=Y(t,x,y,z)$ and $\Ep = \Ep(x,y,z)$ are defined as: 
\beq Y = \frac{R}{\Ep},~~ \Ep = \frac{S}{2}\left[ 1 + \left( \frac{x-P}{S}\right)^2 +\left( \frac{y-Q}{S}\right)^2\right], \label{mtrcfn} \eeq

\noindent with $R = R(t,z)$, $S(z)\geq 0, P(z), Q(z)$ arbitrary functions, and the dust FE yield:

\beq {\dot R}^2 = 2E(z) + {2M(z)}R^{-1};~~
\tilde{M}' = \kappa \tilde{\rho} Y^2 Y',  \label{dust} \eeq
\noindent where a prime and dot denotes differentiation with respect to $z$ and $t$, respectively, $ \tilde{\rho}$ is the energy density, and 
$ \tilde{E} = {2E}\Ep^{-2},~~ \tilde{M} = {M}\Ep^{-3}$ 
are the  energy and mass functions, respectively; also,
we only consider the contracting phase by choosing the negative root of \eqref{dust} below,
and we 
further assume that
$R \geq 0 \text{ and } M\geq 0$.

With $S=1$ and $P = Q=0$, this solution reduces to the LTB solution, while if $R = z \tilde{S}(t)$, $E = E_0 z^2$ with $\tilde{S}$ an arbitrary function, $E_0 =$ constant, $P=Q=0$, and $S = 1$, the FLRW limit is recovered. 
The QSSD model can be regarded as a generalization of the LTB model in which the spheres of constant mass are non-concentric.
The sign of $E(z)$ determines the evolution, and 
all three evolution types can exist in different regions of the same QSSD solution. 
We consider here regions with recollapsing matter ($E< 0$), whence the solution is  \cite{KB2012}:
\beq R = - \frac{M}{2E}(1-\cos \eta),~~ \eta - \sin \eta = \frac{(-2E)^\frac32}{M} (t-t_B(z)), \label{dustsoln} \eeq
\noindent where $t_B(z)$ is an arbitrary function and $\eta(t,z)$ is a parameter.

\subsubsection{Spin-coefficients, scalars and invariant characterization } 

We work with the complex null tetrad, $\{l^a, n^a, m^a {\bar m^a}\}$,  
defined as \cite{szek}:
\beq  \begin{aligned} \ell_a = &\frac{1}{\sqrt{2}}\left( dt +\frac{\Ep {Y'}}{\sqrt{1+2E}} dz\right), n_a = \frac{1}{\sqrt{2}}\left( dt - \frac{\Ep {Y'}}{\sqrt{1+2E}} dz\right),  \\ 
& m_a = \frac{Y}{\sqrt{2}}( dx - i dy),~~ \bar{m}_a = \frac{Y}{\sqrt{2}}( dx + i dy), \end{aligned} \label{nllfrm} \eeq
\normalsize
\noindent and the corresponding frame derivatives are denoted by
$D, \Delta, \delta,~\bar{\delta}$.
The dust condition gives the following coordinate independent relations between the Ricci scalars: $\Phi_{00} = \Phi_{22}  = 2\Phi_{11} = R/4$.

The algebraically independent NP curvature scalars are:

\beq \begin{aligned}  \Phi_{00} =  \frac{\tilde{\rho}}{8 \pi}  &= \frac{2\tilde{M}_{,z}}{Y^2 Y_{,z}},~ \Psi_2 = -\frac{\tilde{M}}{2Y^3}+ \frac{2 \pi}{3}\tilde{\rho},\end{aligned} \label{NPcurv} \eeq

\noindent where, $\pi$ is the standard geometric constant and $\kappa$ is the Einstein gravitational constant. That is,  the Weyl tensor is of algebraic type {\bf D}, and the Ricci tensor is of algebraic type {\bf I}.
The non-zero spin-coefficients $\rho$ and $ \mu$ are given by
\beq \begin{aligned}
\rho = \theta_{(\ell)}&= \frac{1}{\sqrt{2}} \left( \frac{Y_{,t} \Ep - \sqrt{1+2E}}{\Ep Y} \right), \\
\mu = \theta_{(n)} &= -\frac{1}{\sqrt{2}} \left( \frac{Y_{,t} \Ep + \sqrt{1+2E}}{\Ep Y} \right), 
\end{aligned} \label{npend} \eeq
corresponding to the expansion of the two null directions,
and the other non-zero spin-coefficients are $\epsilon,  \tilde{\kappa}$ and $\tilde{\pi}$ and their complex conjugates. We have added a tilde to the expressions for the mass-energy density, $\tilde{\rho}$ and the spin-coefficients $\tilde{\pi}$ and $\tilde{\kappa}$ in order to distinguish from the spin-coefficient, $\rho$, and the constants, $\pi$ and $\kappa$, respectively.

The Cartan-Karlhede algorithm can be used to generate the required set of Cartan invariants for the QSSD spacetime. 
Using the null frame \eqref{nllfrm}, which is the APNF, the zeroth order Cartan-Karlhede algorithm can be applied readily to the Ricci and Weyl tensors. The isotropy group at zeroth order consists of spins. In general, there are two functionally independent non-zero zeroth order Cartan invariants, $\Phi_{00}$ and $\Psi_2$.
At first order, the covariant derivative of the Weyl tensor yields the following algebraically independent quantities:
\beq D\Psi_2,~\Delta \Psi_2, \delta \Psi_2, \bar{\delta}\Psi_2,~\rho, \mu, \kappa, \tau. \label{DWeyl} \eeq

\noindent While from the covariant derivative of the Ricci tensor we find additional quantities:
$D \Phi_{22} + 4 \epsilon \Phi_{22}, \Delta \Phi_{22} - 4 \epsilon \Phi_{22}, \delta \Phi_{22}, \bar{\delta} \Phi_{22}.$
The first order isotropy group is trivial, as spins  affect the form of the spin-coefficients $\kappa, \tau$, and $\epsilon$ along with any quantity differentiated by $\delta$ or its complex conjugate.  
Choosing the frame where $\epsilon$ is real-valued using an appropriate spin, this is now a fixed coframe and any frame derivative of a Cartan invariant is also a Cartan invariant. We are now able to separate the components in equations \eqref{DWeyl} and work with the frame derivatives of $\Phi_{22}$ and $\Psi_2$ and the spin-coefficients directly.
$\epsilon$ and $\tilde{\pi}$ can be chosen as the remaining two functionally independent invariants.
The Cartan-Karlhede algorithm  terminates  at  second order.

The QSSD models admit an AH, defined by the surface $R=2M$, which corresponds to the vanishing expansion of the future-pointing null vector normal to this surface \cite{KB2012}.
To detect the GH, we will consider the covariant derivative of the Weyl tensor. The components of $C_{abcd;e}$ may be expressed in terms of $\Psi_2$, $\Phi_{11}$, $\Delta \Phi_{11}$ and the spin-coefficients.
Using the  algebraic and differential Bianchi identities, the non-zero positive b.w. $+1$ components of $C_{abcd;e}$ in the chosen invariant APNF are \cite{szek}:

\beq C_{1214;3}= C_{1434;3} = C_{1213;4} = C_{1334;4} = 3 \rho \Psi_2, \label{SzkBW10} \eeq

\beq C_{3434;1} = C_{1212;1} = 2C_{1423;1} =  -\frac{2}{3}\Delta  \Phi_{11} -\frac{4}{3}(8 \epsilon + \mu) \Phi_{11}+ \frac{2}{3}\rho(9 \Psi_2+ 2  \Phi_{11})  . \label{SzkBW01} \eeq

The extended Cartan invariant $\rho$ will vanish on the surface $R=2M$. This surface is thus a GH since the extended invariant, $\mu$, which also appears in the covariant derivative of the curvature tensor, is negative within the surface $R=2M$. 
Although this GH coincides with the AH, the geometric interpretation of these two surfaces differs. Relative to the APNF chosen from the Cartan-Karlhede algorithm, it is suggested that for dynamical BH solutions the covariant derivative of the Weyl tensor will be algebraically special on a GH. However, 
it has not yet been shown that this tensor cannot be fully classified using the alignment classification, and hence may not necessarily be of algebraic type {\bf II}.

\paragraph{Examples:} 
In \cite{szek} a simple example of the collapsing QSSD models with no cosmological constant, describing the formation of galactic-sized BHs without any shell-crossings  \cite{KB2012}, was considered.
The AH was determined by
the  vanishing of the extended Cartan invariant $\rho$ (or $\mu$).
A QSSD model generated from a reference LTB solution
was also considered as an example of the formation of a PBH, in which the shell-crossings are contained within the AH. 
The behaviour of the AH was displayed in terms of the zero-sets of $\mu$ or $\rho$ for the expanding and collapsing phases, respectively.
While the family of QSSD BH solutions considered always admit an invariantly characterized AH, only a subset of these solutions permit a physical interpretation as BH solutions due to particular features that can occur during their evolution, such as the appearance of shell-crossing singularities outside of the AH.
The invariant description in terms of Cartan invariants 
can be used to not only give insight into the interpretation of the GH, but also
describe the physical properties of the models in an invariant way \cite{szek}.

\newpage

\subsection{IV: Kastor-Traschen solutions}

Coalescing BHs provide another scenario where DHs appear, and hence provide a test for the GH conjectures. While a numerical simulation of physically realistic coalescing BHs  to examine the algebraic type of the curvature tensor on the horizon is difficult, there are exact solutions that can be studied, such as the Kastor-Traschen (KT) solution \cite{KT}. 

KT found a family of exact closed universe solutions to the Einstein-Maxwell equations with a cosmological constant representing an arbitrary number $N$ of charge-equal-to-mass BHs \cite{KT}. The term
"merger" is used to denote the evolution of initially disjoint trapping horizons which become
a continuous boundary, while "coalescence" denotes the appearance of new
marginal surfaces that enclose the original trapped regions [29]. If coalescence
does not occur, the collision will presumably either produce a naked singularity
(violating the cosmic censorship conjecture) or the dynamics will keep the black
holes apart.
Some aspects of the multiple BH
KT solutions have been investigated \cite{NSH}. 
The KT dynamical two-black-hole solution was discussed in \cite{KT}.
When the sum of the two BH masses is sufficiently small and does not exceed a critical mass $M_C$, the
BHs coalesce and form a larger BH. 
The existence of horizons has been studied in the case of two coalescing BHs \cite{NSH}.
Subsequently,
the existence of invariantly defined quasi-local hypersurfaces 
(characterized by the vanishing
of particular curvature invariants)
in the KT solution 
was studied using GH and SPI in \cite{CMS} and
investigated  
numerically using the frame approach in \cite{CMKT}.

The invariant APNF is constructed from the Cartan-Karlhede algorithm,
and gives rise to scalar Cartan invariants that can be used to determine the geometrical properties of the KT solutions and, in particular, to identify the GH 
hypersurfaces in a foliation independent way. 
These GHs need not be AHs, DHs or MTTs. However, for spherically symmetric dynamical BHs the GH will coincide with the unique DH, $r=2M$. Furthermore, if the sum of the masses of a dynamical KT BH solution are below a certain threshold, it will eventually settle down after merger to a type {\bf D} Reissner-Nordstr{\"o}m-de Sitter BH. This implies that in the quasi-stationary regime there will be a single 3D  GH which coincides with the WIH  \cite{GANG}.

\subsubsection{The APNF in the Cartan-Karlhede Algorithm} 

The KT solution represents $N$ charge-equal-to-mass BHs in a spacetime with a positive cosmological constant, $\bar{\Lambda}$:  
\begin{eqnarray}
& ds^2=-W^{-2}dt^2+W^2(dx^2+dy^2+dz^2)\,;~W=-Ht+\Sigma_{i=1}^N\frac{m_i}{r_i}. \label{KTmetric} & 
\end{eqnarray}
Here $H=\sqrt{\bar{\Lambda}/3}$, where $\bar{\Lambda}\ge0$ is the cosmological constant, $t\in(-\infty,0)$, $m_i$ $i\in [1,N]$ are the BH masses, and $r_i=\sqrt{(x-x_i)^2+(y-y_i)^2+(z-z_i)^2}$, are the BH positions where $r_i = 0$, $i \in[1, N]$, represent a 3D infinite cylinder, with 2D cross-sectional area of $4\pi m_i^2$ for each BH. 
When $N>1$, this solution will generically be of Weyl type {\bf I}  \cite{CMKT}.
If $N=1$ the Weyl tensor is of Weyl type {\bf D} (in which case
the coframe can be fixed entirely at first order \cite{GANG}).

The electromagnetic 4-potential is given by $A=W^{-1}dt$.
The electromagnetic field $F = dA$ is non-null, and a coframe exists such that the energy-momentum tensor is of type {\bf D}. This coframe will be an invariantly defined APNF which can be employed in the Cartan-Karlhede algorithm. To explicitly construct this coframe, we start with  \cite{CMKT}:

\beq t_0 = \frac{dt}{W},~t_1 = W dx,~t_2 = W dy,~t_3 = W dz, \label{KTorthonormal} \eeq

\noindent from which we have the complex null coframe

\beq \begin{aligned} & \ell' = \frac{t_0 - t_1 }{\sqrt{2}},~ n' = \frac{t_0 + t_1 }{\sqrt{2}},~~m' = \frac{t_2 + i t_3}{\sqrt{2}}, \bar{m}' = \frac{t_2 - i t_3}{\sqrt{2}}. \end{aligned} \label{KTnullframe} \eeq
Applying a null rotation about $n$ and then a null rotation about $\ell$ with their respective parameters \cite{kramer} produces a new null coframe $\{\ell, n, m, \bar{m}\}$ for which 
$F^*_{ab} = W^{-2}{\sqrt{W_{,i} W^{,i}}} ( m_{[a}\bar{m}_{b]} - \ell_{[a} n_{b]})$.
The Ricci scalar is $R = 12 H^2$, and relative to this APNF the Ricci tensor,
which is of type {\bf D}, then takes the form:

\beq R_{ab} =  \Phi_{11} (m_{(a}\bar{m}_{b)} + \ell_{(a}n_{b)}). \eeq

By considering the covariant derivative of the Ricci tensor, two real valued
extended Cartan scalar invariant spin-coefficients $\rho$ and $\mu$ can be expressed, using the Bianchi identities, in terms of the components of the Ricci tensor and its covariant derivative:

\beq \rho =  \frac{D \Phi_{11} }{4 \Phi_{11}},~~\mu = \frac{\Delta \Phi_{11}}{4 \Phi_{11}}. \eeq

\noindent Hence, these quantities define the expansion of the outcoming and ingoing null vectors of the invariant coframe:
$\theta_{(\ell)} =  -2\rho, ~~\theta_{(n)} = 2 \mu$ (where $\rho$ and $\mu$ are both real in this case). 

Thus, the APNF can be fully fixed at first order, and  the GH can be identified from the vanishing of the first order Cartan invariants $\rho$ and $\mu$. This is achieved using a direct knowledge of the frame in which $\nabla R$  is of type {\bf{II}}
on the GH. We expect that when $\rho =0 $ or $\mu =0$ that certain positive b.w. components of $\nabla W$ will be zero. For a non-vacuum solution, such as the Kastor-Traschen solution, we can always choose the APNF adapted to the Ricci tensor in a straightforward manner, and it remains to be seen if the APNF adapted to the Weyl tensor will provide a similar characterization of the GH. 

\newpage

\subsubsection{Review of 2 equal mass KT BHs}


In the case of two coalescing BHs, we can adapt coordinates so that the BHs lie on the $x$-axis with both at a coordinate distance $c>0$ from the origin. With this choice, the square root in $W$ becomes $r_{\pm}=\sqrt{(x\pm c)^2+y^2+z^2}$. If the sum of the two BHs masses satisfy the inequality, $M = m_{+} + m_{-} < M_c = \sqrt{3/16\Lambda} = {1}/{4 H},$ 
the BHs will coalesce into a larger single BH, in the sense that a new future outer trapped horizon appears around the BH \cite{NSH}. However, if the sum is greater than this critical mass, the solution is expected to form naked singularities.

In particular, in \cite{CMKT}  we considered the subcritical case $M = 0.5 M_c$ with $H=0.125$, $m_{\pm} = M/2$ and $c = 0.1$, to examine the surfaces where the Cartan invariants $\rho$ and $\mu$ vanish. 
 Within the surfaces defined by $\theta_{(\ell)} =0 $ the other expansion scalar will be negative (i.e., $\theta_{(n)} < 0 $), which is an invariant 3D closed surface defined as a geometrically outer trapped surface (GOTS) in order to distinguish it from the standard GH. Such surfaces will be useful in the descriptions following, and hence the surfaces defined by $\theta_{(n)} = 0$ will also be tracked.  

The GOTSs that surround the collapsing BH  behave in a similar manner to the known foliation-dependent quasi-local horizons with regards to the upper-bound on the total mass \cite{NSH}, in the sense that if the mass of both BHs is below the same critical value, a common GOTSs forms around them as they move together. Similarly, if the sum of the masses are greater than this critical value, there is a strong argument that the outer GOTSs surrounding the BHs will not remain at late times. By tracking the GHs that arise in a dynamical BH solution we can employ them  to determine a smooth, dynamical hypersurface that shields all other horizons and identifies the region of interest.

 We briefly describe the evolution and formation of these GOTSs for this subcritical case \cite{CMKT}.  At early times, the BHs have separate spherical GOTSs surrounding their locations. As the BHs move closer, these GOTSs expand while remaining centred on their respective BH locations while a third GOTS forms around the origin and between the hole of the expanding $\theta_{(n)}=0$ torus centered on the x-axis in the y-z plane. As time increases, the growing GOTSs combine to make a single  GOTS connected through the hole in the $\theta_{(n)}=0$ surface. Within this new outer GOTSs, new spherical GOTSs centred on the BH locations form and expand outwards. 

As time increases further, the outer GOTS grow outward and deform to a torus shaped GOTS aligned along the y-axis, together with a larger GOTS that surrounds all other GOTSs. Moving forward in time, the outermost GOTS expands indefinitely outwards away from the BH and the torus-shaped GOTS defined by $\theta_{(\ell)}=0$ now expands outwards and deforms to create two additional disconnected GOTSs surrounding each of the spherical GOTSs centred on the BHs. These surfaces are contained within a larger GOTS that will also expand outwards away from the locations of the BH. Finally, the intermediate GOTSs merge, creating a new GOTS surrounding the GOTSs centred on the BHs that does not expand away from them. We anticipate that this final state constitutes coalescence and that after this state the spacetime will eventually settle down to a Reissner-Nordstrom-de Sitter BH of mass $m_1+m_2$, where the two GHs then correspond to the bifurcate Killing horizons \cite{GANG}.

\newpage

\section{Two colliding BHs (vacuum)}

Finally, for vacuum spacetimes in which the Ricci tensor is identically zero, perhaps representing the physically interesting situations such as the merger of two BH, the APNF
is chosen so that the  Weyl tensor is explicitly of type {\bf I}. For this choice of frame (which is not the frame utilized in standard computations) we should examine the algebraic properties of the Weyl tensor and its first covariant derivative in this frame. To do this, we must first choose a frame in order to compute the covariant derivatives of the Weyl tensor using the NP spin coefficients. Next, we must verify if $\rho$ and $\mu$  relative to this APNF can be expressed as Cartan invariants. Finally, and especially if starting the simulation at an earlier time before the merger, we must study the covariant derivative of the Weyl tensor in order to better analyize the curvature in the head-on collision.

The aim in the first instance is to apply this to two axisymmetric examples. First, the numerical investigation of axisymmetric 
Kastor-Traschen (KT) 2 BH spacetimes \cite{KT}, which we shall return to later. We note that the KT spacetime is not a vacuum spacetime, but we wish to adapt the APNF to the Weyl tensor in this example.
Second, in the future we will study the axisymmetric  head-on
collision of two unequal mass non-spinning BHs utilizing an APNF. 

\subsubsection{The head-on collision of two unequal mass BHs}

{
To further motivate our investigation, we will first discuss the usage of MOTSs and AHs to study these spacetimes. For theoretical and practical reasons, these surfaces have a more immediate relevance than the EH \cite{Matzner:1995ib,Penrose:1964wq} and can be readily used to extract  quantities of physical interest along with tracking them all the way throughout the BBH merger. In the usual ``pair of pants'' description in the case of two BHs, a spacetime foliation could have four MOTS~\cite{Schnetter:2006yt,2019PRDPook-Kolb}, with an AH as the outer-most of these MOTS.  As noted earlier, the main technical tool for locating MOTSs numerically in previous work is a horizon finder based on the {\em{stability operator}} for MOTS  \cite{AnderssonMarsSimon,Andersson_2009,Pook-Kolb}. Recently, it was shown that, contrary to popular belief, the MOTS stability operator is self-adjoint for a large class of geometric models in GR  \cite{Huber}.

In particular, the problem of numerically
simulating the axisymmetric head-on collision of two unequal mass BHs has been recently considered. In a study \cite{Pook-Kolb}, it was found numerically  that the MOTS associated with the final BH merges with the two (initially disjoint) surfaces associated with the two initial BHs. This produces a ``connected sequence of MOTSs'' which interpolates  between the initial and final state throughout the non-linear BBH
merger process.  Furthermore, the computation was tracked up to and beyond the merger point (see also \cite{Pook}). Lastly, directly following the merger, it was found that MOTS formed which contained self-intersections. 

The fate of the common AH resulting from a BH merger has been studied extensively and is fairly well understood. In general, once a common MOTS forms after a BBH merger, it bifurcates into two surfaces. The outermost of these two surfaces expands and forms the AH for the final BH, while the innermost surface contracts. We note that the original two MOTS that were initially the AHs of the two BHs persist well after the formation of the common horizon \cite{Schnetter:2006yt,2019PRDPook-Kolb}. The world tubes characterized by the two separate MOTSs touch at the instant, $t_{touch}$, after the common horizon has formed, and then penetrate with each other. Of the two surfaces that arise from the bifurcation of the newly formed common horizon, the outer surface  approaches a symmetric equilibrium, while the inner surface distorts and merges with the two interior MOTSs precisely at the instant when they touch.

The fate of the two original AHs during the axisymmetric head-on merger of two non-spinning BHs has been reinvestigated \cite{Pook}, facilitated by a new method for locating MOTSs based on a generalized shooting method in conjunction with a reinterpretation of the stability operator as the analogue of the Jacobi equation for families of MOTSs. The results were subsequently applied to BH mergers, and several such world tubes evolving and connecting through various bifurcations and annihilations were resolved. This produces a consistent picture of the full merger in terms MOTSs. Indeed, the MOTS stability operator consequently provides a natural mechanism to identify MOTSs which should be thought of as BH boundaries
(all MOTSs, other than the two initial ones and the final remnant, lie in the interior and are neither stable nor inner trapped). 
}

\subsubsection{Recent work using scalar invariants}

In an alternative approach to the head-on collision of two unequal mass BHs,
the algebraic properties of the Weyl tensor
through the merger of two non-spinning BHs
was studied numerically  \cite{CPS}. Particular
interest was focused on the conjecture that for such a vacuum spacetime, which
is zeroth-order algebraically general (i.e., of Petrov type
${\bf I}$), a GH, on which the
spacetime is algebraically special and which is identified by the vanishing of a complex scalar invariant (${\mathcal{D}}$), characterizes a smooth foliation
independent surface (horizon) associated with the BH. In the particular
simulation, the level sets of Re(${\mathcal{D}}$) (since Im(${\mathcal{D}}$) = 0) were investigated.
 Now asymptotically (when spacetime is either two separate Kerr BHs or a single
merged BH) in which spacetime is essentially Kerr and of type ${\bf D}$ everywhere, ${\mathcal{D}} \equiv I^3 - 27 J^2$ is zero.
In this case the horizon is conjectured to be located by the fact that the covariant derivative of the Weyl tensor is of type ${\bf II/D}$ there (characterized by the fact that invariants of the covariant derivative of the Weyl tensor are zero). It is these invariants that we also wish to investigate numerically.
In the APNF frame there is some evidence that (in this frame) these surfaces are characterized by the conditions that the spin coefficients $\rho, \mu$ are zero.

The ultimate goal is the  study of curvature invariants in a {\em non-axisymmetric} binary BH merger \cite{PCS}. 
Recently, we have studied numerically the GH conjecture by tracing the level zero set of the magnitude of the complex scalar polynomial invariant
of the Riemann tensor, $|{\mathcal{D}}|$, through a quasi-circular merger of two non-spinning, equal mass BHs (by approximating the level--0 sets of ${\mathcal{D}}$).
The numerical results presented provide evidence that a
(unique) smooth GH can be identified throughout all stages
of the binary BH merger \cite{PCS}. To study this more comprehensively in future we need to compute the covariant derivative of the Weyl tensor and the spin coefficients 
$\rho, \mu$ in an APNF (ab initio, since the null frame used in previous work is not  a preferred invariant frame).

\newpage
\section{Characterizing the APNF}

In order to define the APNF, we wish to completely fix the frame in an appropriate invariant way
so that the resulting quantities are Cartan scalar invariants. We need to consider the covariant derivative of the Weyl tensor. We are especially interested in the spin coefficient $\rho$, which is related to the 
positive b.w. terms in the covariant derivative of the Weyl tensor, and whose real part is minus the expansion of the outcoming null vector $k$ of the invariant coframe. Indeed, in the chosen
APNF the vanishing of $\rho$ will be important in identifying the GH. We will follow the definitions and
notation of \cite{kramer}, and refer to eqns. therein by the prefix ES (e.g. eqn. (ES7.1) is eqn. (7.1) in  \cite{kramer}).  We note that the notation used in this section for the null directions differs from that in \cite{kramer}; the null direction $\ell$ in their notation corresponds to $n$ while $k$ corresponds to $\ell$ in this paper.

First, we apply a (Lorentz transformation) null rotation about $n$ (to fix $E$) to set:

\beq
\Psi_0 =0, \label{APNFa}
\eeq 
and a spin (to fix $\theta$) so that $\Psi_1$ is real
\beq
Im(\Psi_1) = 0 \label{APNFb}. 
\eeq 
In this frame we note that ${\mathcal{D}} \equiv I^3 - 27 J^2 = {\Psi_1}^2 F(\Psi_1,\Psi_2,\Psi_3,\Psi_4$) where F is a quartic polynomial function, so that in the APNF the type {\bf{II/D}} condition becomes
$\Psi_1 = 0$ (one real condition).

Equations \eqref{APNFa} and \eqref{APNFb} are the principle conditions satisfied by the APNF. We need to completely fix the frame invariantly. We shall do this in one of two ways. In the second (see section 5 below) we shall fix the frame in a manner suitable for numerical computations. In the first we define a frame more suitable for theoretical considerations.  In the first version of the APNF, we shall  apply a  null rotation about $\ell$ (to fix $B$)
to set $\Psi_3 =0$ (or, alternatively, to set the  spin coefficient $\sigma$ to zero --  see later),
and a boost ($A$) to set either the real or complex part of the spin coefficient 
$\epsilon$ to zero. This completely fixes the APNF invariantly, but we note that we expect the
GH to be characterized by the vanishing of $\Psi_1$ and $\rho$, the zeros of which are not affected by the choice of $B$ and  $A$. For reference we note that in this frame $\Psi_1$ is of b.w. +1, the directional covariant derivatives of the Weyl spinor $D\Psi_1, \delta \Psi_1, \bar{\delta}\Psi_1$ are of b.w. +2,+1,+1, respectively (and $D\Psi_2$ is of b.w. +1), and the spin coefficient  $\kappa$   is of b.w. +2 and the spin coefficients $\rho, \sigma, \epsilon$ are all of b.w. +1.

\subsection{The Geometric Horizon}

We shall use the notation of $\hat{X}$ to denote the value of a quantity ${X}$ on the GH; i.e., we have that 
\beq
\hat{\Psi}_1= 0. \label{psi}
\eeq
We expect the GH to be a 3D surface, so that at a point on the GH the directional derivatives
of ${\Psi}_1$ tangent to the GH should also be zero there. In general, we can choose the spatial derivatives (spanned by $\bf{m}$ and $\bar{\bf {m}}$) $\delta \Psi_1$ and $\bar{\delta}\Psi_1$ to be zero at such a point (and along the GH) by a (Lorentz) null rotation (affecting $\bf{m}$ and $\bar{\bf {m}}$).
However, we might anticipate that as well as $\hat{\Psi}_1= 0$ (and $\hat{\rho}= 0$), the quantities
$\widehat{D 
 \Psi}_1, \widehat{\delta  \Psi}_1, \widehat{{\bar{\delta}} \Psi}_1$ (and $\widehat{D \rho}, \widehat{\delta \rho}, \widehat{\bar{\delta}  \rho}$, as well as $\hat{\kappa}, \hat{\sigma}$) might vanish on the GH, so that affecting
Lorentz transformations on a neighbourhood ${\mathcal{U}}$ of the GH may be problematic.

If we assume that the GH is {\em{null}}, then $\hat{\nabla} \hat{\Psi}_1$ is null and 
\beq
|\hat{\nabla \Psi}_1|^2 = 0,
\eeq
so that either (1) $ \ell.\widehat{\nabla \Psi}_1=0$ and hence $ \widehat{D \Psi}_1 = 0$ (i.e., $\ell$ is orthogonal to the GH), or (2) $ \ell.\widehat{\nabla \Psi}_1 \neq 0$ (and by continuity $ {D}{\Psi}_1\neq 0$ on ${\mathcal{U}}$). If the initial separation between the two black holes is small enough, the MOTSs are approximately null surfaces and hence isolated horizons.

Let us consider case (2) first. Here $ {D} {\Psi}_1 \neq 0$ on ${\mathcal{U}}$. We can then apply a null rotation to set  ${\delta \Psi}_1 = {\bar{\delta}\Psi}_1 = 0$  on ${\mathcal{U}}$. Eqn. (ES7.32a) then implies that
$\kappa \neq 0$ on ${\mathcal{U}}$, and hence we can apply a Lorentz transformation to set $\sigma = 0$ on ${\mathcal{U}}$.
Since the GH is null, this then implies that  $ \widehat{\delta \Psi}_1 = 0$ (and $ \widehat{\Delta \Psi}_1 = 0$). It then follows from eqns.
(ES7.32a,b,e) that as well as $\widehat{\delta \Psi}_1=0$ and $\hat{\sigma} = 0$ and  $ \widehat{D \Psi}_1 \neq 0$  (b.w. +2 term) we have that $ \widehat{D \Psi}_2 \neq 0$ (b.w. +1 term), which then leads to a contradiction. (We note that there is one special case in which $\Psi_4 = 0$ and we can effectively interchange $\ell$ and $n$ and we are in case (1)).

\subsection{Null surface}

We shall henceforward concentrate on case (1) in which  
\beq 
\widehat{D \Psi}_1 = 0.\label{dpsi}
\eeq
We shall also explicitly choose our remaining null rotation (i.e., $B$) such that $\Psi_3=0$. The remaining boost freedom does not affect the zeros of, for example, $\rho$, but we shall fix the frame freedom (i.e., $A$) so that $Re(\epsilon)=0; (\epsilon+ \Bar{\epsilon}) = 0$.

Since 
\beq \ell.\widehat{\nabla \Psi}_1=0,
\eeq we have that 
\beq
\widehat{\delta \Psi}_1 . \widehat{\bar{\delta}  \Psi}_1 = 0,
\eeq 
and since $\Psi_1$ is real,  hence
\beq
\widehat{\delta \Psi}_1 =0, ~~ \widehat{{\bar{\delta}}  \Psi}_1 = 0. \label{delpsi}
\eeq 
Note that in this case ${\Delta \Psi}_1 \neq 0$ and that $\nabla \Psi_1  = {\Delta \Psi}_1 \ell$.
We want to show that $\hat{\rho}=0$.

The Bianchi identities (ES7.32a,b,e) immediately yield
\beq
\hat{\kappa}=0, ~~ \hat{\sigma}=0,\label{scs}
\eeq
and that $\widehat{D \Psi}_2 = 3 \hat{\rho} \hat{\Psi}_2$. The Ricci identities (ES7.21a,b,c) then give
$$ \widehat{D \rho}  - \widehat{\bar{\delta}} {\hat{\kappa}} = \hat{\rho}^2 $$
$$ \widehat{D \sigma}  - \widehat{\delta  \kappa} = 0  $$
$$ \widehat{\delta \rho}  -   \widehat{\bar{\delta}} \hat{\sigma} = \hat{\rho}( \hat{\bar{\alpha}} + \hat{\beta} ) + (\hat{\rho} - \hat{\bar{\rho}} )\hat{\tau}$$
and the eqns. (ES7.21 c,p,q) and (ES7.32f,g,h) and the commutation relations (ES7.6).  These eqns. are compatible with
$\hat{\rho}=0, ~\widehat{D \rho} =0, ~
\widehat{\delta \rho} =0, ~ \widehat{\bar{\delta} \rho} = 0$  characterizing the GH.  Let us make this plausibility argument more convincing.

In the final choice of our invariant APNF we can choose $B$ so that either $\Psi_3=0$ or $\sigma=0$
on  ${\mathcal{U}}$, and by a boost we can choose $\epsilon$ to be real or complex. As we noted earlier, 
we do not expect the zeros of, for example, $\Psi_1$ on ${\mathcal{U}}$ to depend upon these choices, and in computations we will take $\Psi_3=0$ for concreteness (and computational ease). But in order to further justify our choice of APNF and make the argument firmer, let us consider the case where we choose our frame in which $\sigma=0$
on  ${\mathcal{U}}$.

Since from (\ref{scs}) we have that 
$\hat{\kappa}=0$ and $ \hat{\sigma}=0$, before proceeding we must first consider the case in which
$\kappa$ vanishes on  ${\mathcal{U}}$ (since in this case we cannot use a null rotation to set 
$\sigma = 0$  on  ${\mathcal{U}}$). We also choose $\epsilon$ to be real (and $\Psi_1$ and $D\Psi_1$
are real). When $\kappa=0$, we immediately find from (ES7.32a) and (ES7.21a) that $2\rho + \epsilon$ is real and hence $\rho$ is real, and hence from the other Ricci identities that 
\beq
D\rho = \rho^2 + \sigma \bar{\sigma}, D\sigma = 2\rho\sigma,
\eeq
and hence 
\beq
D^2\rho = 6\rho D\rho - 4\rho^3.
\eeq
We look for smooth solutions for $\rho$ in terms of a parameter $v$ orthogonal to the GH and for which
the GH is specified by $v=0$ (i.e., $\hat{v}=0$); namely, we look for power law solutions in $v$ to this DE in the vicinity of $v=0$. One such series solution is not analytic at $v=0$, so we consider the second solution. But this solution leads to $\rho=0$ on  ${\mathcal{U}}$, which is not consistent with the existence of a GH, and hence leads to a contradiction.

Hence we have that $\kappa \neq 0$ on  ${\mathcal{U}}$, and hence we can choose our frame so that
\beq
\sigma = 0
\eeq
on  ${\mathcal{U}}$, which we assume hereafter. [Note that although $\kappa$ is non-zero on ${\mathcal{U}}$, we have that  $\hat{\kappa}=0$ and so we can only use a null rotation using $B$ to set
$\sigma =0$ on ${\mathcal{U}}/GH$. But since $\hat{\sigma}=0$, we can find a smooth solution in which 
$\sigma =0$ on ${\mathcal{U}}$.] We shall also assume $\epsilon$ is complex here ($\epsilon+\bar{\epsilon} =0$).

Eqns. (ES7.21) then yield
\beq
\widehat{D \rho}= \hat{\rho}^2,~~  \widehat{\delta \kappa}=0,    
\eeq
and $\widehat{\delta \bar{\kappa}}=0$, and $\widehat{\delta \rho}$ is given by eqn. (ES7.21k).
Then using the commutation relation (ES7.6b) operating on $\rho$, and using eqns. (ES7.21) (and especially (ES7.21c,d,e)), on the GH we find that
\beq
- \hat{\rho} \hat{C}     + (\hat{\rho}  - \hat{\bar{{\rho}}})[(\hat{\rho}  + \hat{\bar{\rho}}) - (\hat{\epsilon}  - \hat{\bar{\epsilon}})] \hat{\tau} =0,      
\eeq
where $$C \equiv [ -(\bar{\epsilon}\beta+\bar{\beta}{\epsilon})  - (\bar{\epsilon}\alpha+\bar{\alpha}{\epsilon}) + (\bar{\epsilon}-{\epsilon})\alpha  + \alpha\rho - \bar{\alpha}\bar{\rho} +  
({\pi}+\bar{\pi}) {\epsilon} - \pi({\rho} + {\bar{\rho}})].$$
This eqn. implies that the real part of $\rho$ vanishes on the GH.

\section{Axisymmetric example}

To study BHs (with axisymmetry) we define the APNF precisely according to the following invariant conditions:
\beq
\Psi_0 =0, ~~Im(\Psi_1) = 0, ~~\Psi_4 =0 \text{ and } \Psi_3 = e^{i\psi},
\eeq 
 \noindent where $\psi$ is  some function representing the phase of $\Psi_3$.  This alternative (second) frame is more suitable for numerical computations. In particular, $\Psi_4 = 0$ can be implemented over an entire region of a generic Petrov type {\bf I} spacetime, whereas $\Psi_3$ may vanish at particular points and this will introduce complexities into the numerical procedure in fixing the frame. The final condition is invariantly defined since $|\Psi_3| =1$ fixes the boost parameter to a particular value.

We wish to implement this frame choice numerically for BBH mergers, both in the general case and the special axisymmetric case. In some axisymmetric examples this can be partially implemented analytically.

\subsubsection{Weyl adapted APNF for the 2-mass KT solution}

We start from the coframe in eqn. \eqref{KTnullframe} for the 2 equal mass KT metric with $r_{\pm} = \sqrt{(x\pm c)^2 +y^2 + z^2 }$ in the subcritical case (where explicitly, $m_{\pm} = \frac{M_c}{4}$, and $H = 0.125$ and $c=0.1$).  We will employ a null rotation about $n$ in order to set $\tilde{\Psi}_0 = 0$. Relative to the frame in eqn. \eqref{KTnullframe}, $\Psi_0 \neq 0$, and so the result of a null rotation about $n$ with the complex-valued parameter, $E$, to make $\ell$ a principal null direction (PND) is:

\beq \tilde{\Psi}_0 =  E^4 \pd + 4 E^3 \pc + 6E^2 \pb + 4 E \pa + \po = 0. \label{PND3} \eeq 

\noindent We will follow the notation defined in \cite{Nerozzi} in order to solve for the aligned null direction $\ell$. To simplify this polynomial, we will introduce the reduced variable  
\beq z = \pd E + \pc, \label{PNDzb} \eeq
\noindent to give a new equation to solve:
\beq z^4 + 6 {H} {z}^2 +4 {G}{z} + {K} = 0, \label{PND4} \eeq
where the quantities are defined by
\beq \begin{aligned} {H} &= \pd \pb - \pc^3 \\ 
{G} &= \pd^2 \pa - 3\pd \pc \pb + 2\pc^3 \\ 
{K} &= \pd^2 I - 3 H^2. \end{aligned} \eeq

\noindent In the definition of ${K}$ we have used the SPI defined in \eqref{Iinv}. 

The solution for the PNDs can be expressed using three new quantities:
\beq \begin{aligned} {\alpha'}^2 & = 2\pd \lambda_1 - 4{H}, \\
{\beta'}^2 & = 2\pd \lambda_2 - 4{H}, \\
{\gamma'}^2 & = 2\pd \lambda_3 - 4{H}, \end{aligned} \label{allgreek2} \eeq

\noindent where the $\lambda_i$ variables are the eigenvalues of the specific matrix $Q$ built from the Weyl scalars:
\beq \begin{aligned} \lambda_1 & = \left(P + \frac{1}{3P} \right), \\
\lambda_2 &= -\left( e^{\frac23 \pi i} P + e^{\frac43 \pi i } \frac{I}{3P} \right),\\ 
\lambda_3 &= -\left( e^{\frac43 \pi i} P + e^{\frac23 \pi i } \frac{I}{3P} \right), \end{aligned} \label{WeylEV} \eeq
with 
\beq P \equiv \left[ J + \sqrt{ J^2 - (I/3)^3} \right]^{\frac13},\label{WEVP}  \eeq
\noindent $I$ and $J$ are defined in eqns. \eqref{Iinv} and \eqref{Jinv}, and the $\Psi_i$ are given explicitly (in the initial frame \eqref{KTini}) by the following expressions: 

\beq \begin{aligned} \Psi_0 &= \bar{\Psi}_4 = \frac{(W W_{,z,z} - W W_{,y,y} -3 W_{,z}^2 + 3 W_{,y}^2) + i(6 W_{,z} W_{,y} - 2 W_{,y,z} )}{2 W^4} \\ 
\Psi_1 &= - \bar{\Psi}_3 = \frac{(3 W_{,x} W_{,y} - W_{,x,y} W)+ i (3W_{,x} W_{,z} - W W_{,x,z}  )}{2 W^4} \\
\Psi_2 &= \frac{(W W_{,z,z} + W W_{,y,y}  - 2 W W_{,x,x} - 3 W_{,z}^2 - 3 W_{,y}^2 + 6 W_{,x}^2}{6 W^4}.  \end{aligned} \label{KTini} \eeq

\noindent We can explicitly write $P$ in terms of $W$ and its derivatives.

The solution to \eqref{PND4} is then 

\beq \begin{aligned} z_1 &= \frac12 ({\alpha'} + {\beta'} + {\gamma'}), \\
z_2 &= \frac12 ({\alpha'} -  {\beta'} - {\gamma'}), \\
z_3 &= \frac12 (-{\alpha'} + {\beta'} - {\gamma'}), \\
z_4 &= \frac12 (-{\alpha'} - {\beta'} + {\gamma'}), \end{aligned} \label{PNDzsol} \eeq
\noindent and solutions to the original equation in \eqref{PND3} can be found using \eqref{PNDzb}; e.g.,

\beq E = \frac{z_1- \Psi_3}{\Psi_4} = \frac{\alpha' + \beta' + \gamma'- 2\Psi_3}{2\Psi_4}. \eeq

\noindent Relative to this new frame, the non-zero Weyl scalars are $\tilde{\Psi}_1, \tilde{\Psi}_2, \tilde{\Psi}_3$ and $\tilde{\Psi}_4$. The choice of the root $z_i$ is equivalent to a discrete Lorentz transformation between possible PNDs. As a simple choice to fix the frame, we will choose to set $\Psi_4= 0$ by applying a null rotation about $\ell$ and solving the resulting third degree polynomial:

\beq \tilde{\Psi}_4 + 4 B \tilde{\Psi}_3 + 6B^2 \tilde{\Psi}_2 + 4 B^3 \Psi_1 = 0. \label{APNFbkt} \eeq

\noindent Finally, we will choose the spin parameter, $\theta$, so that $\Psi_1$ is real-valued (i.e., $Im(\Psi_1) = 0$) and fix the boost parameter, $A$, so that $|\Psi_3|^2 = 1$. 

We have now fixed the APNF completely. We can explicitly write down the expressions for $E,B, \theta$ and $A$ and hence specify all of the remaining Cartan scalars. We can then investigate the zeros of $\Psi_1$ and $\rho$, which we shall do in future work. In the current KT example some of these computations can be done analytically. For example, by computing the Weyl scalars for the initial frame \eqref{KTini}. The parameters $B$ and $E$ can be computed explicitly and the relevant quantities relative to the APNF written down. To illustrate this, we will compute the frame where $\ell$ is now a PND of the Weyl tensor after a null rotation with parameter $E$ given by eqns. \eqref{PNDzb} and \eqref{PNDzsol}; the non-zero Weyl scalars are then:
\beq \begin{aligned}  \tilde{\Psi}_0 & = 0, \\
\tilde{\Psi}_1 & = \Psi_1 + 3 E \Psi_2 + 3 E^2 \Psi_3 + E^3 \Psi_4, \\
\tilde{\Psi}_2 &= \Psi_2 + 2 E \Psi_1 + E^2 \Psi_4, \\
\tilde{\Psi}_3 & = \Psi_3 + E \Psi_4, \\ 
\tilde{\Psi}_4 &= \Psi_4. \end{aligned} \eeq
\noindent By applying a spin, we can then set $\tilde{\Psi}_1$ to be {\it real-valued} with the parameter

\beq \theta = \frac{1}{4i} \ln \left( \frac{\bar{\tilde{\Psi}}_1}{\tilde{\Psi}_1} \right), \eeq

\noindent and so 

\beq \tilde{\tilde{\Psi}}_1  = \sqrt{\tilde{\Psi}_1 \tilde{\bar{\Psi}}_1} = |\tilde{\Psi}_1|. \eeq

To continue, we can apply a null rotation about $\ell$ with parameter $B$ which satisfies equation \eqref{APNFbkt}. Relative to this new frame a boost can always be chosen so that $|\Psi_3|^2 = 1$. 

It remains to outline a numerical scheme for implementing the above procedure, which will then produce a numerical solution for $\Psi_1$ and $\rho$ relative to this APNF, which can then be used for studying the merger of two BHs. We will present the results of this analysis in a future paper.
\newpage

\section*{Acknowledgements}  
 
This work was supported through NSERC (A. A. C.). 

\end{document}